\documentclass[aps,prc,two column,showpacs,amsmath,amssymb,nofootinbib]{revtex4}
\usepackage {lscape}
\usepackage{graphicx}
\usepackage{amsmath}

\usepackage[bookmarks=true]{hyperref}
\usepackage{color}

\hypersetup{
	colorlinks=true,
	pdfborder={0 0 0},
	citecolor=blue,
	linkcolor=blue,
	urlcolor=blue
}

\begin{document}

\title{Gamow-Teller transition strengths for selected  ${fp}$ shell nuclei}

\author{Vikas Kumar$^{1}$\footnote{Electronic address: vikasphysicsiitr@gmail.com},  Praveen C.~Srivastava$^2$\footnote{Electronic address: praveen.srivastava@ph.iitr.ac.in} and Anil Kumar$^{2}$}
\address{$^{1}$Department of Physics, Central University of Kashmir, Ganderbal - 191201, INDIA \\
$^{2}$Department of Physics, Indian Institute of Technology, Roorkee 247 667, 
INDIA }

\date{\hfill \today}
\begin{abstract}

 We have reported a systematic shell model description of the experimental Gamow-Teller transition strength for
 $^{44}$Sc $\rightarrow$ $^{44}$Ca,  $^{45}$Ti $\rightarrow$ $^{45}$Sc, $^{48}$Ti $\rightarrow$ $^{48}$V,
$^{66}$Co $\rightarrow$ $^{66}$Ni, and
$^{66}$Fe $\rightarrow$ $^{66}$Co transitions using KB3G and GXPF1A
interactions for $fp$ model space. In order to see the importance of higher orbital for $^{66}$Co $\rightarrow$ $^{66}$Ni and 
$^{66}$Fe $\rightarrow$ $^{66}$Co transitions, we have reported the shell model results with $fpg_{9/2}$ space using GXPF1Br+$V_{MU}$ interaction.
We have obtained the qualitative agreement for the individual transitions, while the calculated summed transition
strengths closely reproduce the observed ones.

\end{abstract}
\maketitle

\section{Introduction}
 The Gamow-Teller ${(GT)}$ transition is a nuclear week interaction process which is used as basic input to study the structure of atomic nuclei
\cite{Bohr,suhonen,Osterfeld,Rubio,Fujita,Harakeh,Broglia,Fujita1,Bai,ser,archana,kumar_epja}.
To estimate electron-capture (EC) reaction rates in the case of $\beta^+$ decay, we need Gamow-Teller ($GT$)  strength [$B(GT)$] distributions. 
 The EC reactions on medium-mass nuclei play a significant role in astrophysical phenomena such as
core-collapse (type-II) supernovae (SNe); thermonuclear type (type Ia) SNe; heating and cooling processes in crusts of accreating neutron stars.
Thus to understand these process it is highly desirable to precisely calculate $GT$ strengths using suitable nuclear models.
Experimental $GT$ strengths can be obtained from $\beta$-decay and charge-exchange reactions. The $\beta$-decay measurements are limited to small $Q$-value window,
while with charge exchange reactions like ($p$, $n$), ($^2$He, $d$) and ($^3$He, $t$) are useful tools to study the relative values of $B(GT)$ strengths up to high excitation energies.
The Gamow-Teller transition study for the $^{48}$Ti($^{3}$He,$t$)$^{48}$V reaction is reported in Ref. \cite{ganioglu_48ti48v},
the highly fragmented $GT$ strength distributions for $^{48}$Ti are observed in this experiment. 
The experimental $GT$-strengths corresponding to $^{66}$Co $\rightarrow$ $^{66}$Ni and $^{66}$Fe $\rightarrow$ $^{66}$Co transitions are available in 
Ref. \cite{66co66ni66fe66co}.
Theoretical investigation to study strong magnetic dipole ($M1$) transitions and $GT$ strengths for
$fp$ shell nuclei reported in Refs. \cite{LZ1,LZ2,talmi,martinez1,richtler,kota,caurier,mar}.

In the present work our aim is to calculate the $GT$ strengths and to compare the theoretical 
results with the experimental data. Also, we have calculated $GT$ strengths distributions 
at higher excitation energies. This might be very useful for upcoming experimental data.
It is also possible to predict half-lives using $GT$ strengths as an input.

In the present work, we have performed shell model calculations to obtained the $GT$-strengths for 
$^{44}$Sc $\rightarrow$ $^{44}$Ca, 
$^{45}$Ti $\rightarrow$ $^{45}$Sc
and $^{48}$Ti $\rightarrow$ $^{48}$V transitions using GXPF1A and KB3G effective interactions in the full $fp$ model space.
We have also reported the $GT$-strength results corresponding to $^{66}$Co $\rightarrow$ $^{66}$Ni and 
$^{66}$Fe $\rightarrow$ $^{66}$Co transitions using $fp$ and $fpg_{9/2}$ spaces. 
In the Table 1, we have given list of $fp$ shell nuclei considered in the present work for $GT$-strength calculations, 
the number of $GT$ transitions, transitions up to the excitation energy in MeV and the references are 
given in the last column for comparison with the theoretical results.

\begin{table*}
\begin{center}
\caption{\label{tab:summary1} This table shows initial and final nuclei, 
the number of $GT$ transitions, transitions up to the excitation energy in MeV and the references are 
given in the last column for comparison with the theoretical results.}
\begin{tabular}{cccccccccc}
\hline
\hline
\\
 S.No. & Initial   & Final & Transitions (No.) &    EXPT.    & GXPF1A  &  KB3G  &    GXPF1Br+$V_{MU}$  &   Ref.  \\
\hline  \\  
 1. & $^{44}$Sc($2^{+}$) & $^{44}$Ca($1^{+}$,$2^{+}$,$3^{+}$) &  50   &  3.301 &   19.204  & 16.883 & -& \cite{NDS_2011}\\
 2.&$^{45}$Ti($\frac{7}{2}^{-}$) & $^{45}$Sc($\frac{5}{2}^{-}$,$\frac{7}{2}^{-}$,$\frac{9}{2}^{-}$) & 50  & 1.662 & 10.022 &  9.384 &  -& \cite{NDS_2008}\\   
 3.&$^{48}$Ti($0^{+}$) & $^{48}$V($1^{+}$) &  350  & 12.646 &  13.048 &  12.983 & - &\cite{ganioglu_48ti48v}\\  
 4.&$^{66}$Co($1^{+}$) & $^{66}$Ni($0^{+}$,$1^{+}$,$2^{+}$) & 100\footnote{For GXPF1Br+$V_{MU}$ interaction we have calculated 300 eigen values.}& 3.752   & 15.506 & 19.540 & 18.730 &\cite{66co66ni66fe66co} \\
 5. & $^{66}$Fe($0^{+}$) & $^{66}$Co($1^{+}$) & 100\footnote{For GXPF1Br+$V_{MU}$ interaction we have calculated 300 eigen values.}  &   2.236   &  13.546      &  17.880 &  13.638 & \cite{66co66ni66fe66co} 
 
\\
\hline   
\hline
\end{tabular}
\end{center}
\end{table*}
\section{Details of the Shell model calculation}
 The shell-model effective Hamiltonian can be express in terms of single-particle energies and two-body matrix elements,
\begin{eqnarray}
        \nonumber H&=&\sum_{\alpha}\varepsilon_{\alpha}{\hat N}_{\alpha} \\
        &&+ \frac{1}{4}\sum_{\alpha\beta\delta\gamma JT}\langle  
j_{\alpha}j_{\beta}|V|j_{\gamma}j_{\delta}\rangle_{JT}A^{\dag}_{JT;j_{\alpha}j_{\beta}}A_{JT;j_{\delta}j_{\gamma}},
        \label{eqa:1}
 \end{eqnarray}
    where, $\alpha=\{nljt\}$ is the single-particle orbitals and $\varepsilon_{\alpha}$ is corresponding to the single-particle
energies.\\
 $\hat{N}_{\alpha}=\sum_{j_z,t_z}a_{\alpha,j_z,t_z}^{\dag}a_{\alpha,j_z,t_z}$ is the particle number operator.
  $A_{JT}$
 and $A_{JT}^{\dag}$) are the fermion pair annihilation and creation
operator, respectively.
 $\langle
j_{\alpha}j_{\beta}|V|j_{\gamma}j_{\delta}\rangle_{JT}$ are the
two-body matrix elements coupled to spin $J$ and isospin $T$.

 To obtain the $GT$-strengths we have performed shell model calculations in the $fp$ model space using the 
 KB3G \cite{KB3G} and GXPF1A\cite{GXPF1A} interactions.
 In order to see the importance of higher orbital for $^{66}$Co $\rightarrow$ $^{66}$Ni and $^{66}$Fe $\rightarrow$ $^{66}$Co 
 transitions we have also included the results with $fpg_{9/2}$ space using the GXPF1Br+$V_{MU}$ interaction \cite{gxpf1br}.
 Although, GXPF1Br+$V_{MU}$ interaction is for $fpg_{9/2}d_{5/2}$ space, but here we are not allowing protons/neutrons to occupy in the
 $d_{5/2}$ orbital. For $^{66}$Co $\rightarrow$ $^{66}$Ni and $^{66}$Fe $\rightarrow$ $^{66}$Co transitions
we fix minimum six particles in the $f_{7/2}$ orbital, while for both protons and neutrons we 
allow maximum 2 neutrons in the $g_{9/2}$ orbital. Thus we put same truncations for both protons and neutrons.
 The shell model calculations are performed using the code NuShellX@MSU\cite{NUSHELLX}. 

The Gamow-Teller strength  $B(GT)$ is calculated using the following expression,
\begin{equation}
 {B(GT_{\pm})} = \frac{1}{2J_i + 1} f_q^2 \, |{\langle {f}|| \sum_{k}{\sigma^k\tau_{\pm}^k} ||i \rangle}|^2,
\end{equation}
where  $\tau_+|p\rangle = |n\rangle$ , $\tau_-|n\rangle  = |p\rangle$, $f_q$ is the quenching factor,
the index $k$ runs over the single particle orbitals, 
$|i \rangle$ and $|f \rangle$ describe the state of the parent and daughter nuclei, respectively. 
The reported $B(GT)$ and summed $B(GT)$ values 
are quenched by a quenching factor $q=0.66$ \cite{vikas}.

It is possible to improve further $GT$ strengths results by adding the effect of two-body currents (2BCs) in the quenching factor 
\cite{M.Bender,T.R.Rodriguez,J.Menendez,A.Ekstrom}. Recently for the $ab ~initio$ calculations it was suggested that if 
we use evolve operator then there is no need to use quenching factor \cite{nature2}.





\begin{figure}
\resizebox{0.45\textwidth}{0.35\textwidth}{\includegraphics{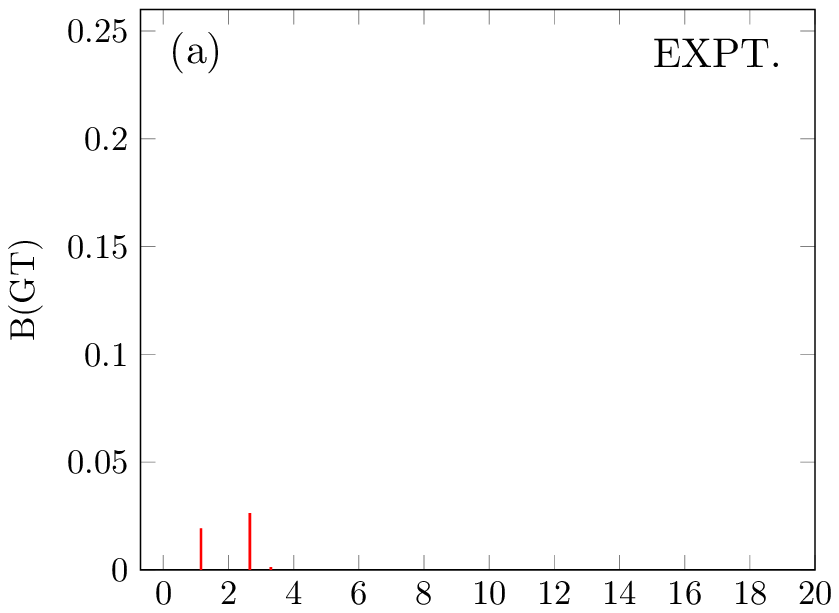}}
\resizebox{0.45\textwidth}{0.35\textwidth}{\includegraphics{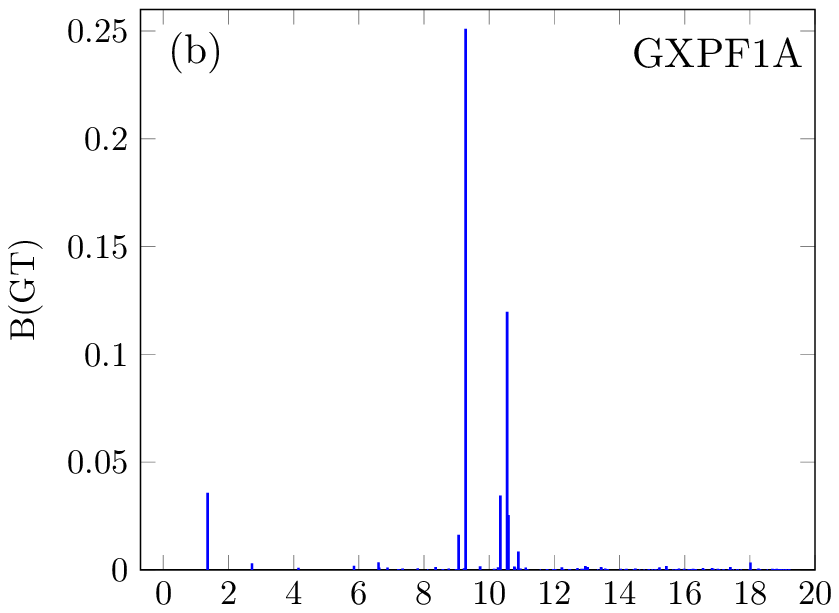}}
\resizebox{0.45\textwidth}{0.35\textwidth}{\includegraphics{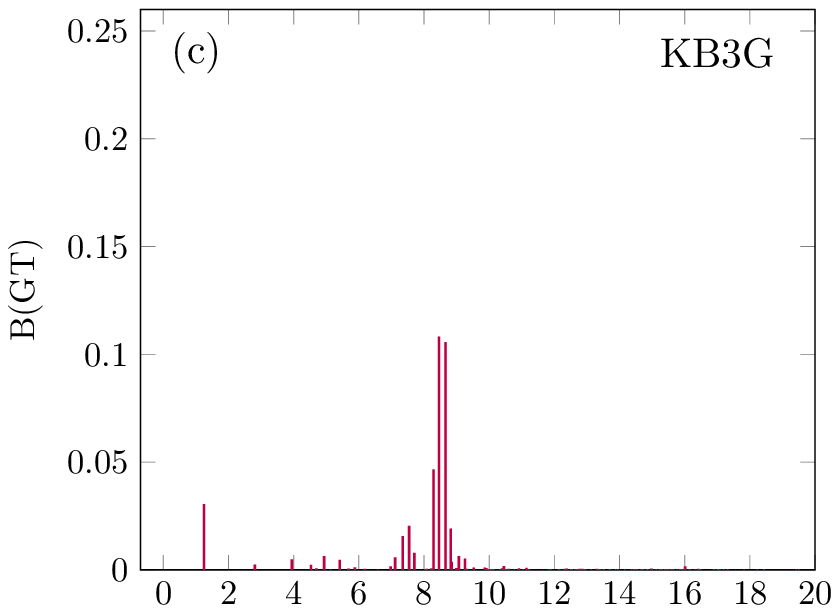}}
\resizebox{0.45\textwidth}{0.35\textwidth}{\includegraphics{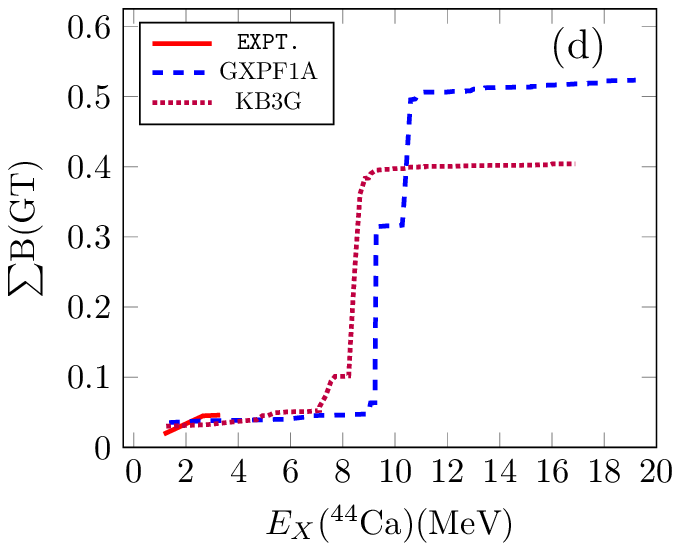}}
\caption{\label{fig6} Comparison of experimental and theoretical $B(GT)$ distributions for $^{44}${Sc}.}
\end{figure}

\section{Comparison of theoretical and experimental $B(GT)$ strength distributions}
The comparison between calculated and experimental $GT$-strengths distributions 
for different transitions are reported below.

\subsection{\bf$^{44}$Sc $\rightarrow$ $^{44}$Ca}

 Fig.~\ref{fig6} shows the experimental and calculated shell model $B(GT)$ strength distributions for the transition $^{44}$Sc $\rightarrow$ $^{44}$Ca. 
The $B(GT)$ values from ${2^+}$ ground state of $^{44}$Sc($2^{+}$)$\rightarrow$ $^{44}$Ca$({1^+, 2^+, 3^+})$ states have been calculated without any truncation. Fig.~\ref{fig6}(a) represents the experimental data observed through the $\beta^{+}$-decay $^{44}$Sc$\rightarrow ^{44}$Ca\cite{NDS_2011}, 
Fig.~\ref{fig6}(b), represents the shell-model calculation using the GXPF1A interaction, 
Fig.~\ref{fig6}(c), the shell-model calculation using the KB3G interaction, 
and Fig.~\ref{fig6}(d), the running sums of $B(GT)$ as function of excitation energy $E_x(^{44}$Ca).
The first $B(GT)$ strength observed at 1.157 MeV are predicted by both the shell model calculations but the second one at 2.657 MeV is predicted slightly smaller than the experiment in both the shell model calculations. The third observed $B(GT)$ strength at 3.301 MeV is predicted by both the shell model calculations at 4.148 and 3.947 Mev, respectively. From the sums of $B(GT)$ strength figure it is clear that both the shell model results are in good agreement with the observed summed $B(GT)$ strengths, it indicates that the $fp$ space is able to produce the observed results 
in the case of $^{44}$Sc $\rightarrow$ $^{44}$Ca transition at low excitation energies. Both the shell model calculations predict a large 
number of $B(GT)$ values at $\sim$ 10 MeV which are not yet observed in the experiment. 

\subsection{\bf$^{45}$Ti $\rightarrow$ $^{45}$Sc}

\begin{figure}
\resizebox{0.45\textwidth}{0.35\textwidth}{\includegraphics{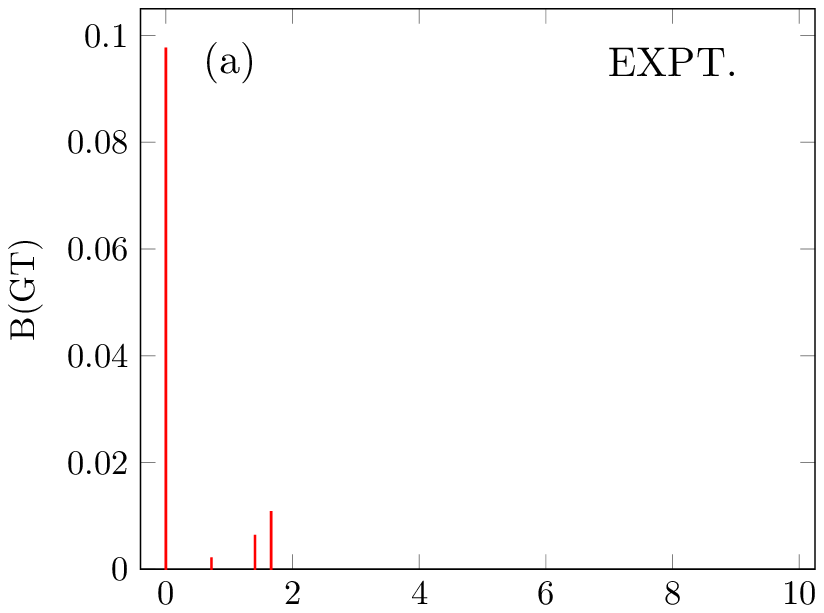}}
\resizebox{0.45\textwidth}{0.35\textwidth}{\includegraphics{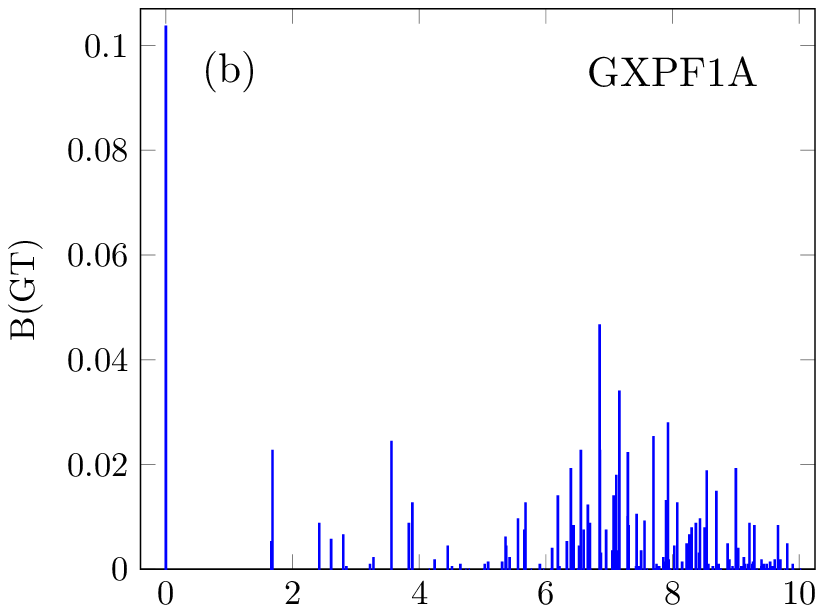}}
\resizebox{0.45\textwidth}{0.35\textwidth}{\includegraphics{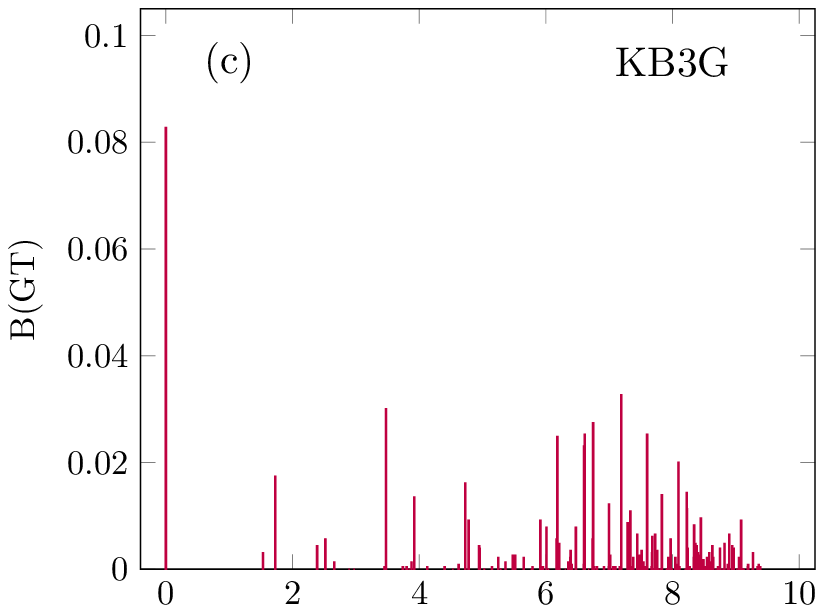}}
\resizebox{0.45\textwidth}{0.35\textwidth}{\includegraphics{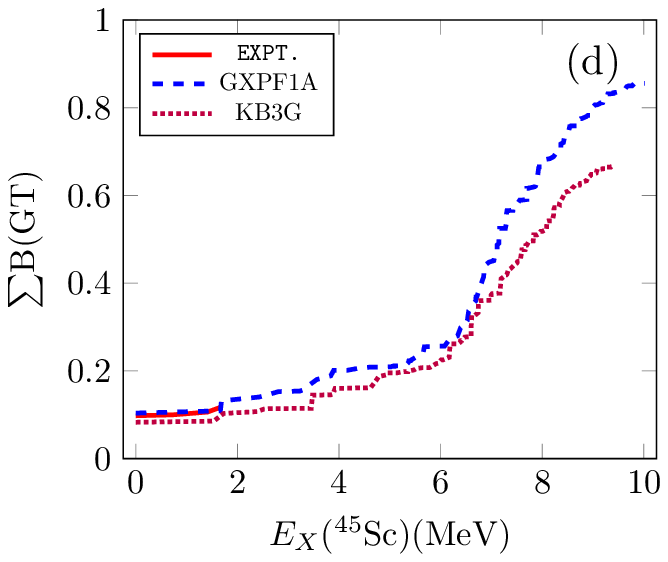}}
\caption{\label{fig3} Comparison of experimental and theoretical $B(GT)$ distributions for $^{45}$Ti.}
\end{figure}

 Fig.~\ref{fig3} displays a comparison between the shell-model calculations and the experimental $GT$ strength distribution for the transition
$^{45}$Ti $\rightarrow$ $^{45}$Sc. 
We have calculated $B(GT)$ values from ground state of $^{45}$Ti ($\frac{7}{2}^{-}$) $\rightarrow$ $^{45}$Sc($\frac{5}{2}^{-}$,$\frac{7}{2}^{-}$,$\frac{9}{2}^{-}$) states without any truncation.
Fig.~\ref{fig3}(a) presents the experimental data observed through the $\beta^{+}$-decay \cite{NDS_2008}.
Fig.~\ref{fig3}(b) depicts the shell-model calculation using the GXPF1A interaction, 
Fig.~\ref{fig3}(c), the shell-model calculation using the KB3G interaction, 
and Fig.~\ref{fig3}(d), the running sums of $B(GT)$ as a function of the excitation energy $E_x(^{45}$Sc). 
There are four $B(GT)$ transition strengths observed in the experiment at 0, 0.72, 1.408, and 1.662 MeV lies between 0.002 - 0.011, these low lying $B(GT)$ strengths are successfully produced by both the shell model calculations. Both the shell model calculations predict the highly fragmented $GT$ strengths at excitation energies $E_x(^{45}$Sc) $\sim$ 2-10 MeV which are not observed in the experiment.
The concentrated $GT$ strengths predicted by the theory at higher excitation energies may be observed in the future experiments. 
The calculated shell model results for
the sum of $B(GT)$ strengths at lowest energy states are in good agreement with the experiment and the trend 
of both the shell model results are following the same pattern at higher excitation energies. 

\subsection{\bf$^{48}$Ti $\rightarrow$ $^{48}$V}

\begin{figure}
\resizebox{0.45\textwidth}{0.35\textwidth}{\includegraphics{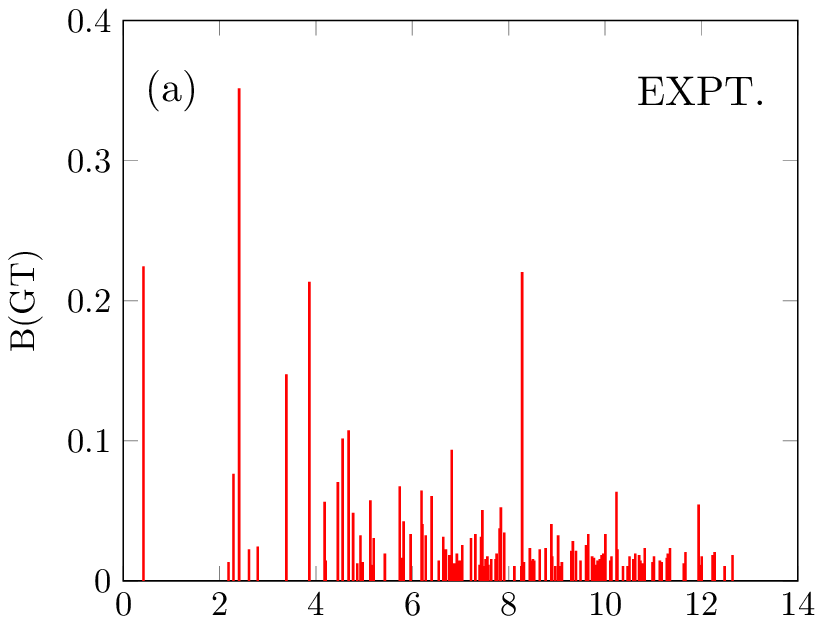}}
\resizebox{0.45\textwidth}{0.35\textwidth}{\includegraphics{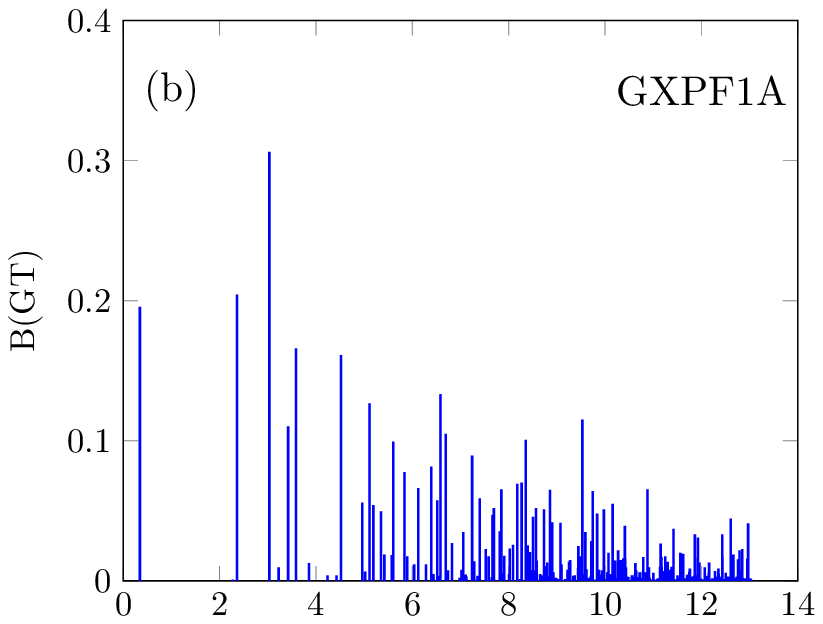}}
\resizebox{0.45\textwidth}{0.35\textwidth}{\includegraphics{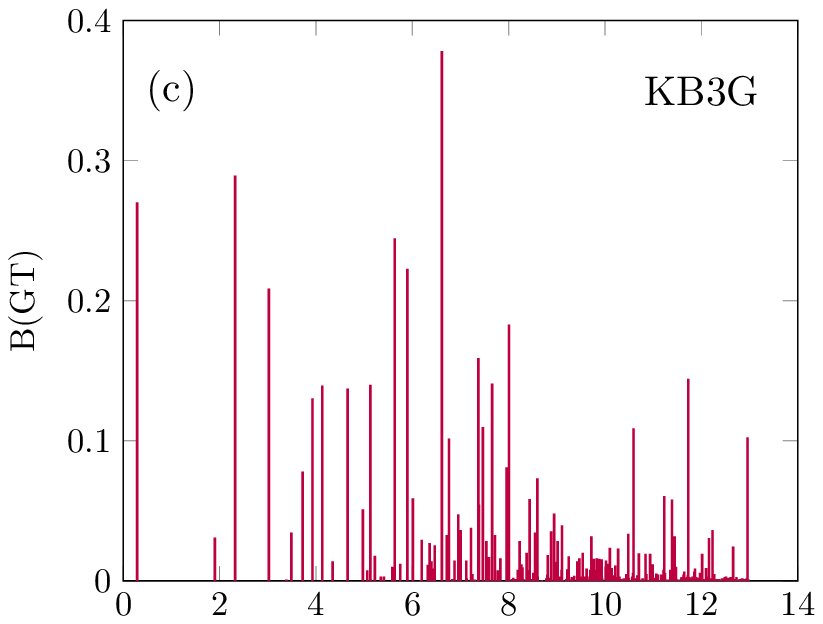}}
\resizebox{0.45\textwidth}{0.35\textwidth}{\includegraphics{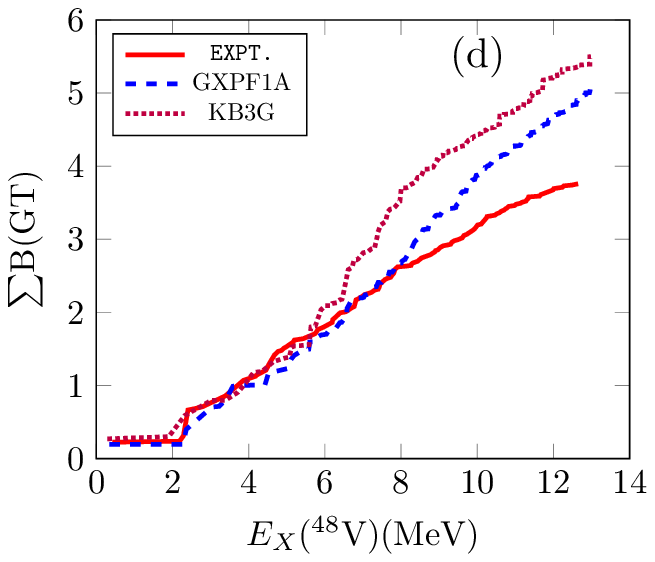}}
\caption{\label{fig2} Comparison of experimental and theoretical $B(GT)$ distributions for $^{48}${Ti}.}
\end{figure}

 Fig.~\ref{fig2} displays a comparison between the shell-model calculations and the experimental $B(GT)$ strength distribution for the transition 
$^{48}$Ti $\rightarrow$ $^{48}$V. 
We have calculated $B(GT)$ values from ground state of $^{48}$Ti$(0^+ )$ to $^{48}$V$(1^+ )$ states without any truncation.
Fig.~\ref{fig2}(a) presents the experimental data observed through the charge-exchange reaction 
$^{48}$Ti($^3${He},$t$)$^{48}$V \cite{E.Ganioglu},
Fig.~\ref{fig2}(b) depicts the shell-model calculation using the GXPF1A interaction, Fig.~\ref{fig2}(c), the shell-model calculation using the KB3G interaction, and Fig.~\ref{fig2}(d), the running sums of $B(GT)$ as a function of the excitation energy $E_x(^{48}$V). 
Fig.~\ref{fig2}(a), shows that the $GT$ strength is highly fragmented and distributed over many discrete states, the same pattern is also 
predicted from both the shell model calculations. The four dominated $GT$ values ranges from 0.147 to 0.351 observed for the transitions from the 
$J^{\pi} = 0^{+}$, ground state of $^{48}$Ti to the $1^+$ states of $^{48}$V at excitation energies $E_x$ = 0.421, 2.406, 3.387, and 3.864 MeV. The calculated shell model intensities for these transitions are similar to the measured ones. At higher excitation energies, both the shell model calculations predict some more dominated transitions which are not observed in the experiment, while one dominated $GT$ strength observed in the experiment at $E_x$ = 3.387 MeV is missing in both the calculations.
The GXPF1A interaction generated an excitation energy closer to the experimental one than the energy obtained employing the KB3G interaction. From fig.~\ref{fig2}(d) the 
summed $B(GT)$ strength plot, the summed $B(GT)$ strength predicted by the GXPF1A interaction is more closer to the experiment than the KB3G interaction.
The summed $B(GT)$ strength by KB3G is in agreement with the experiment at lower excitation energy but not at higher excitation energy,
overall,  the summed $B(GT)$ strength predicted by GXPF1A interaction matched with observed ones better than KB3G.


\begin{figure*}
\begin{center}
\resizebox{0.45\textwidth}{0.35\textwidth}{\includegraphics{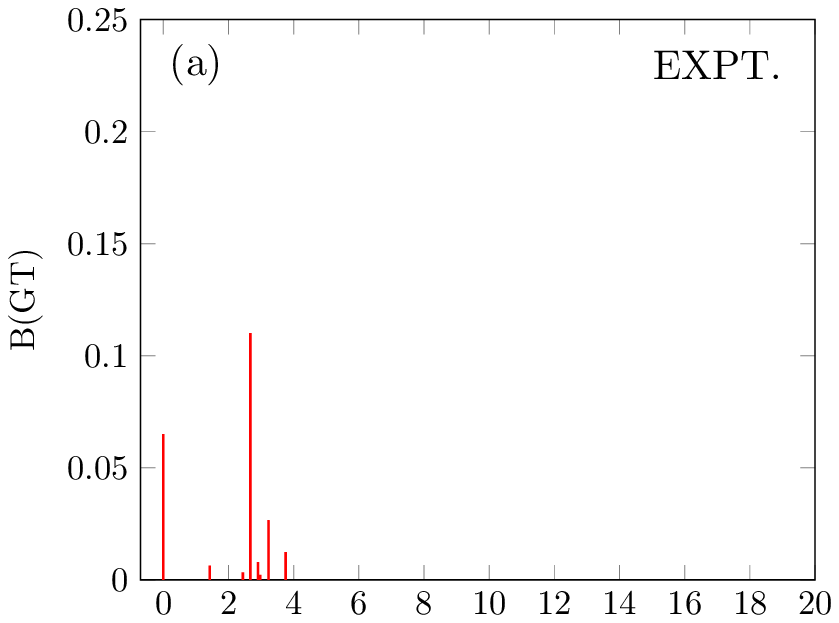}}
\resizebox{0.45\textwidth}{0.35\textwidth}{\includegraphics{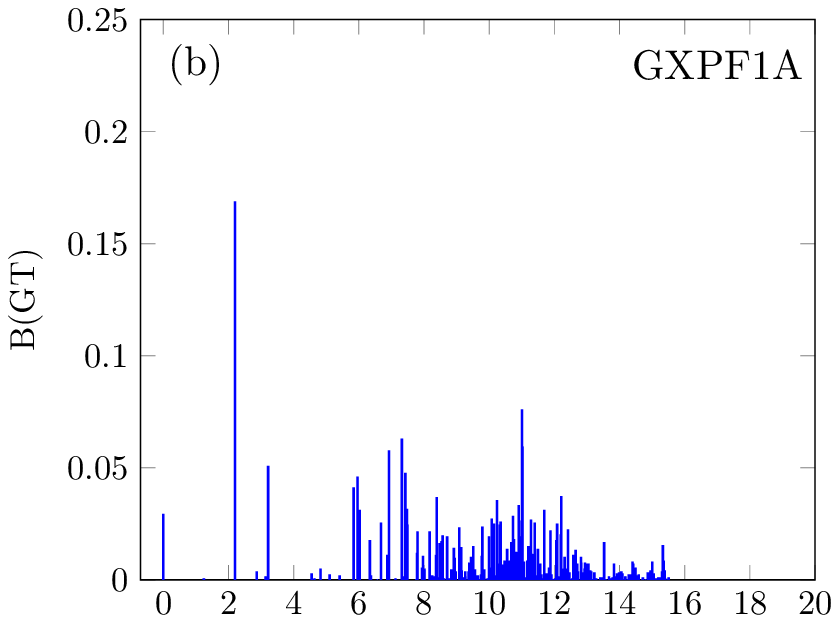}}
\resizebox{0.45\textwidth}{0.35\textwidth}{\includegraphics{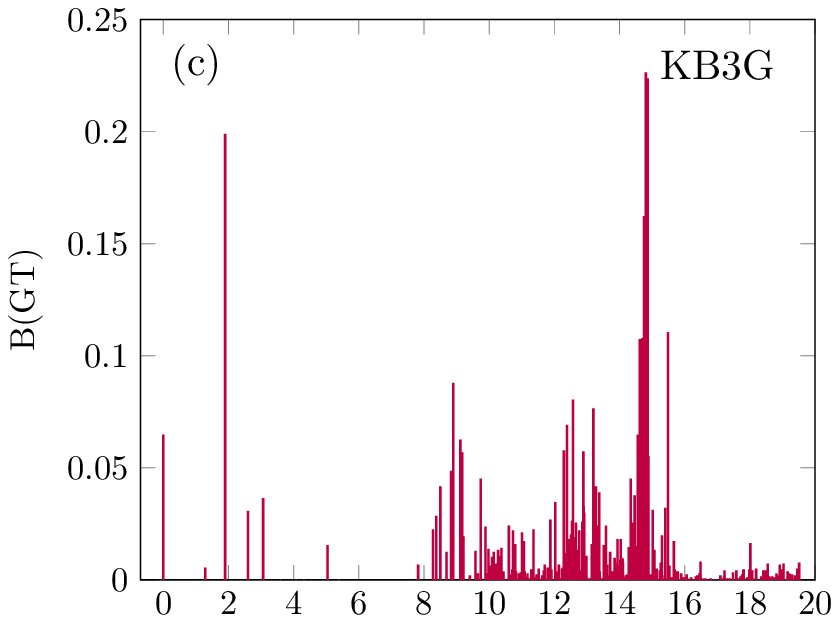}}
\resizebox{0.45\textwidth}{0.35\textwidth}{\includegraphics{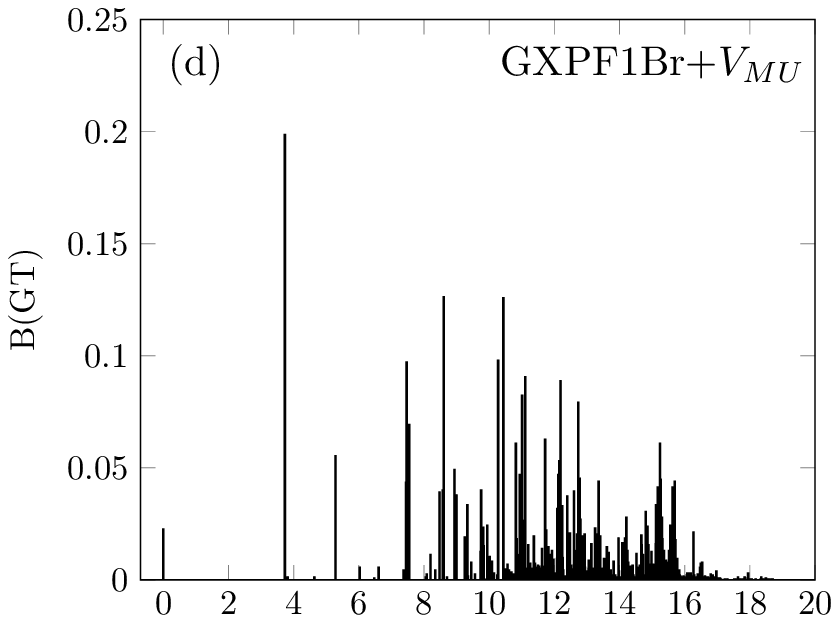}}
\resizebox{0.45\textwidth}{0.35\textwidth}{\includegraphics{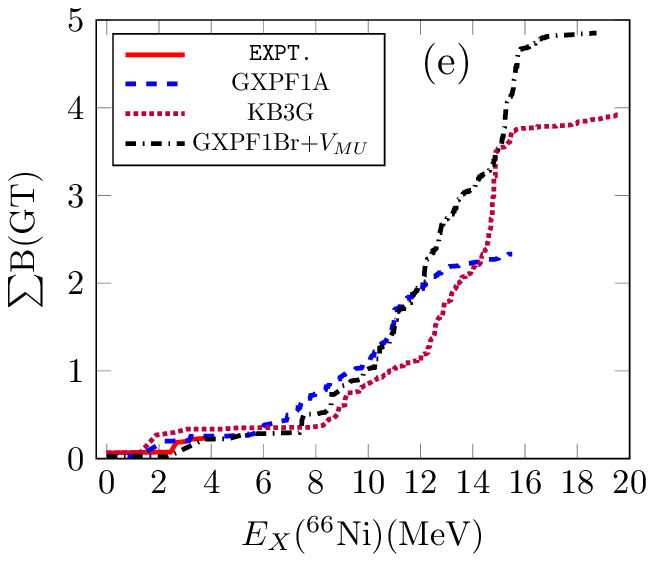}}
\caption{\label{fig4} Comparison of experimental and theoretical $B(GT)$ distributions for $^{66}${Co}.}
\end{center}  
\end{figure*}

\begin{figure*}
\begin{center}
\resizebox{0.45\textwidth}{0.35\textwidth}{\includegraphics{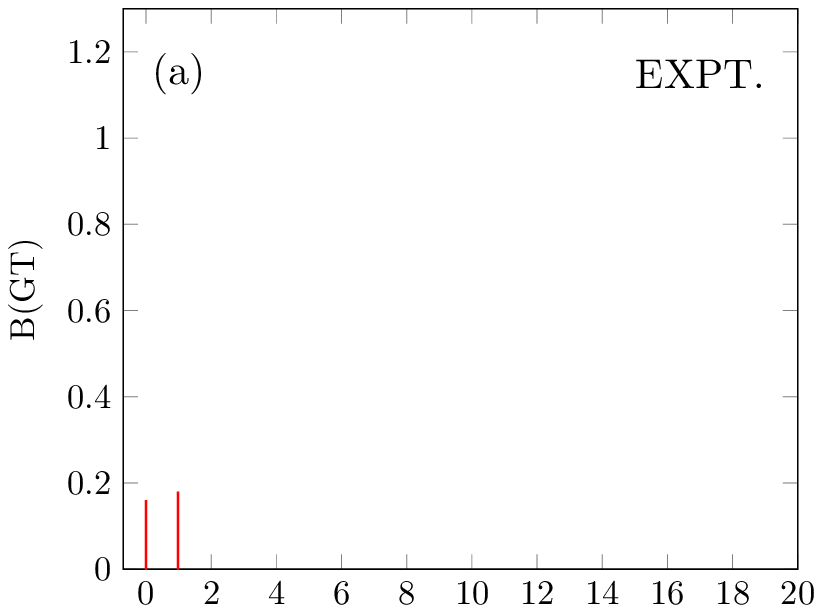}}
\resizebox{0.45\textwidth}{0.35\textwidth}{\includegraphics{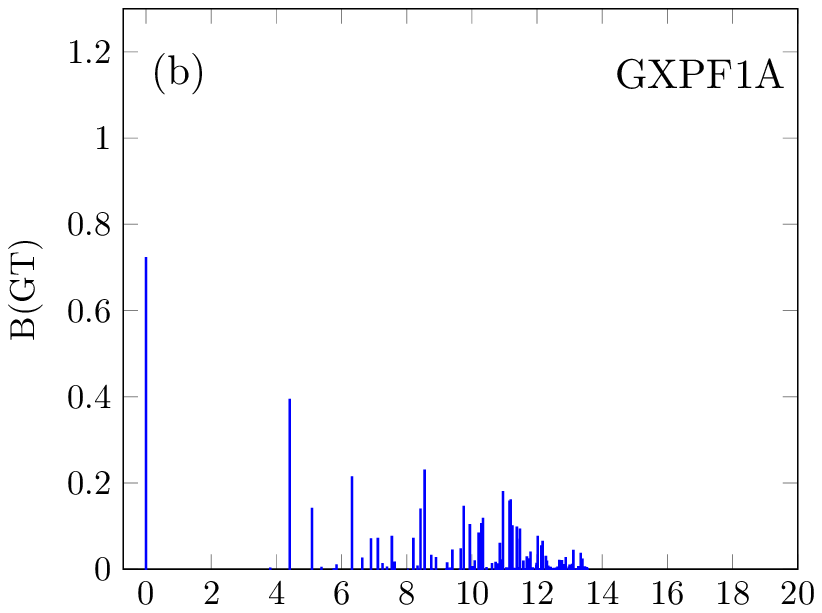}}
\resizebox{0.45\textwidth}{0.35\textwidth}{\includegraphics{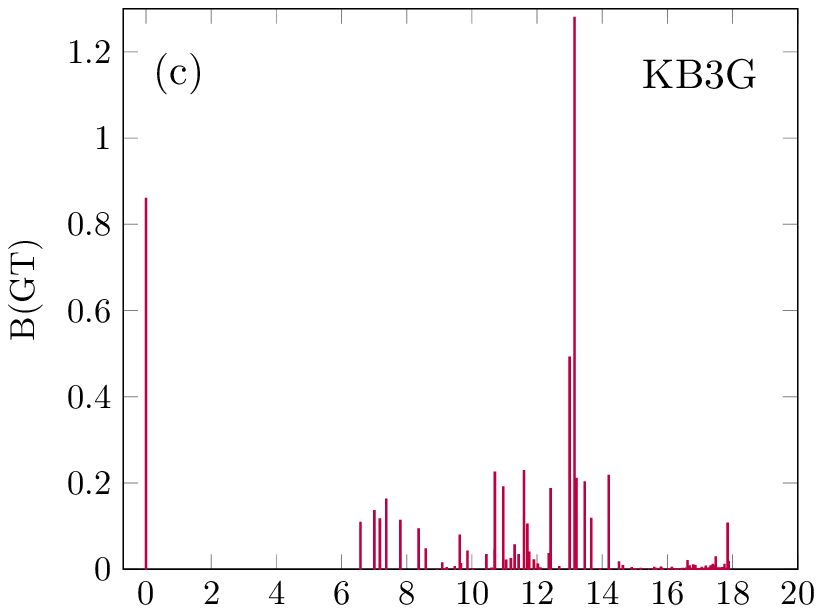}}
\resizebox{0.45\textwidth}{0.35\textwidth}{\includegraphics{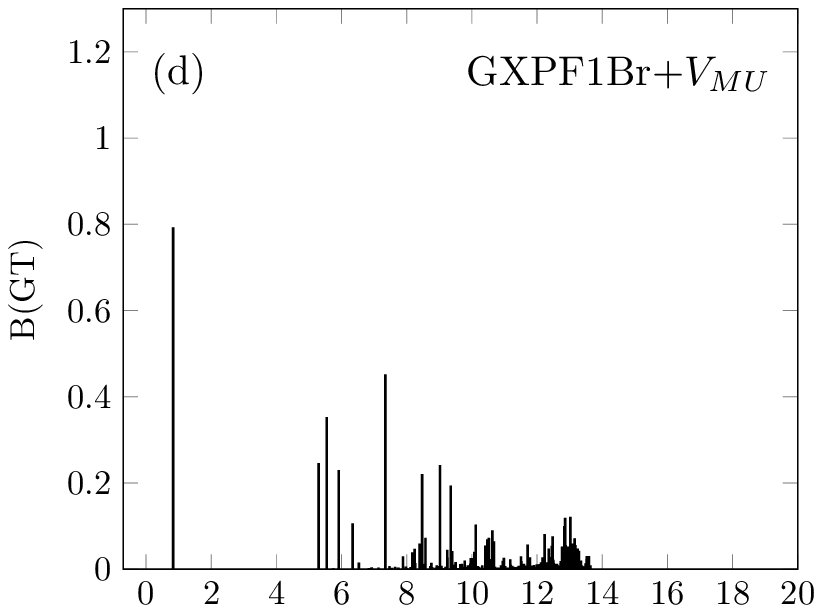}}
\resizebox{0.45\textwidth}{0.35\textwidth}{\includegraphics{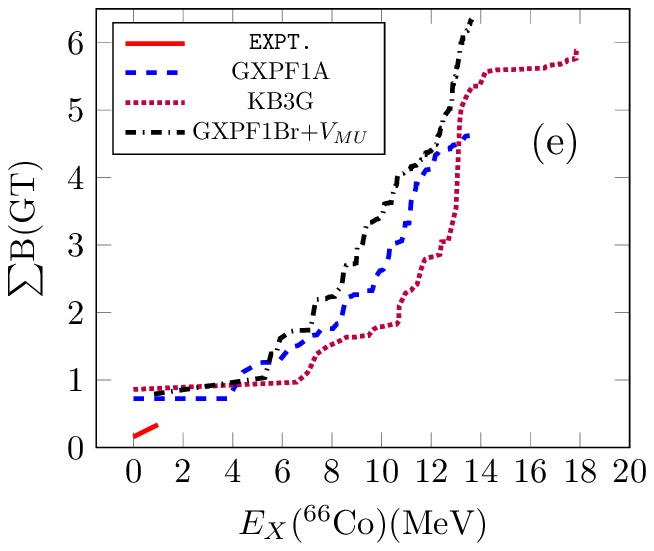}}
\caption{\label{fig5} Comparison of experimental and theoretical $B(GT)$ distributions for $^{66}${Fe}.}
\end{center}  
\end{figure*}

\subsection{\bf$^{66}$Co $\rightarrow$ $^{66}$Ni}

Fig.~\ref{fig4} shows the experimental and calculated shell model $B(GT)$ strength distributions for the transition $^{66}$Co $\rightarrow$ $^{66}$Ni. 
We have calculated $B(GT)$ values from ground state of $^{66}$Co($1^{+}$) $\rightarrow$ $^{66}$Ni($0^{+}$,$1^{+}$,$2^{+}$) states without any truncation using GXPF1A and KB3G interactions. Fig.~\ref{fig4}(a) represents the experimental data observed through the $\beta^{-}$-decay $^{66}$Co$\rightarrow ^{66}$Ni\cite{M.Stryjczyk}, Fig.~\ref{fig4}(b), represents the shell-model calculation using the GXPF1A interaction, 
Fig.~\ref{fig4}(c), the shell-model calculation using the KB3G interaction, 
Fig.~\ref{fig4}(d), represents the shell-model calculation using the 
GXPF1Br+$V_{MU}$ interaction for $fpg_{9/2}$ model space, and Fig.~\ref{fig4}(e), the running sums of $B(GT)$ as function of excitation energy $E_x(^{66}$Ni). 
There are eight $GT$ transition strengths which are observed from the ground state of $^{66}$Co to different excited states of
$^{66}$Ni at 0, 1.425, 2.443, 2.671, 2.907, 2.974, 3.228, and 3.752 MeV, overall all these eight $GT$ transition strengths are 
also produced in all the shell model calculations. All the interactions predict several weakly excited states with $B(GT)$ values for example: the GXPF1A with $B(GT)$ values of 0.001-0.174 in the range 4.00-15.506 MeV, the KB3G with $B(GT)$ values of 0.001-0.226 in the range 4.00-19.540 MeV and the GXPF1Br+$V_{MU}$ with $B(GT)$ values of 0.001-0.127 in the range 5.000-18.730 MeV. All these predicted weakly excited states are not observed in the experiment. Overall the sum of $B(GT)$ strengths predicted by GXPF1A interaction is more closer to experiment than other two interactions. 

\subsection{\bf$^{66}$Fe$\rightarrow$ $^{66}$Co} 
The shell-model calculations and the experimental $GT$ strength distributions for the transition
$^{66}$Fe $\rightarrow$ $^{66}$Co are presented in the Fig.~\ref{fig5}.
We have calculated $B(GT)$ values from ground state of $^{66}$Fe($0^{+}$) $\rightarrow$ $^{66}$Co($1^{+}$) states without any truncation using GXPF1A and KB3G interactions.
The experimental data observed through the $\beta^{-}$-decay $^{66}$Fe$\rightarrow ^{66}$Co\cite{M.Stryjczyk} 
are shown in  Fig.~\ref{fig5}(a), in Fig.~\ref{fig5}(b), the shell-model calculation using the GXPF1A interaction, 
in Fig.~\ref{fig5}(c), the shell-model calculation using the KB3G interaction, 
in Fig.~\ref{fig5}(d), the shell-model calculation using the GXPF1Br+$V_{MU}$ interaction, 
in Fig.~\ref{fig5}(e), the running sums of $B(GT)$ as function of 
the excitation energy $E_x(^{66}$Co). Two dominant $GT$ transition strengths are observed in the experiment 
from $^{66}$Fe$(0^+)$ $\rightarrow$ $^{66}$Co$(1^+)$ states at $E_x(^{66}$Co) = 0 and 0.982 MeV, first experimental
$GT$ transition strength is predicted in both GXPF1A and KB3G shell model calculations while in GXPF1Br+$V_{MU}$ interaction the first $B(GT)$ value is shifted to the higher excitation energy. In all the shell model calculations, the second observed $B(GT)$ value is missing. It is 
found that the GXPF1A interaction generated an excitation energy and $B(GT)$ strengths more closer to the experiment 
than the KB3G interaction. The shell model calculations predict several  excited states with small $B(GT)$ values
in the 4 - 14 MeV region in GXPF1A, 6.5 - 17.8 MeV region in KB3G and 5.298 - 13.638 MeV region in GXPF1Br+$V_{MU}$ effective interaction. These several weakly $GT$ transitions strengths are not observed in the experiment, these theoretical results may be serve as the input for the future 
experiments. 

The results of $GT$ strengths with GXPF1A and KB3G are different, this might be due to different originality
           of these two interactions. The GXPF1A interaction is developed from G-matrix with state-of-art fitting procedures, while KB3G interaction is a monopole-corrected version of KB3 effective interaction.
There are also differences in the single-particle energies adopted. Overall the density of states are generally dense
for the GXPF1A interaction.

\section{Summary and Conclusion}
In the present work we have reported shell model result in the $fp$ model space for recently measured $GT$-strengths of 
 $^{44}$Sc $\rightarrow$ $^{44}$Ca,  $^{45}$Ti $\rightarrow$ $^{45}$Sc,
$^{48}$Ti $\rightarrow$ $^{48}$V,
$^{66}$Co $\rightarrow$ $^{66}$Ni, and
$^{66}$Fe $\rightarrow$ $^{66}$Co transitions.
To see the importance of $g_{9/2}$ orbital, we have perform calculation in $fpg_{9/2}$ space for 
 $^{66}$Co $\rightarrow$ $^{66}$Ni and
$^{66}$Fe $\rightarrow$ $^{66}$Co transitions. 
The qualitative agreement is obtained for the individual $B(GT)$ transitions,
while the calculated summed transition strengths closely reproduce the observed ones. 
In the case of 
$^{48}$Ti $\rightarrow$ $^{48}$V, $^{45}$Ti $\rightarrow$ $^{45}$Sc, $^{66}$Co $\rightarrow$ $^{66}$Ni
and $^{66}$Fe $\rightarrow$ $^{66}$Co transitions, theoretical strengths are larger than the experimental value. 
Thus further experimental results are needed for these strengths.
Results of the present work will add more information to earlier works.\\

\section*{Acknowledgments}
 V.K. acknowledges financial support from Research Project C.U.K. Ganderbal-191201
and a research grant from SERB (India), EEQ/2019/000084.
We acknowledge Prayag computational
facility at Physics Department, IIT-Roorkee.


\end{document}